\documentclass[aps,10pt,prd,superscriptaddress,showpacs,nofootinbib,twocolumn]{revtex4-2}
\usepackage{amssymb}
\usepackage{graphicx}
\usepackage{makeidx}
\usepackage{amsfonts}
\usepackage{amsthm}
\usepackage{amsmath}
\usepackage{eurosym}
\usepackage{hyperref}
\usepackage{revsymb}

\usepackage{tensor}

\hypersetup{
    colorlinks = true,
    linkcolor = blue,
    anchorcolor = blue,
    citecolor = blue,
    filecolor = blue,
    urlcolor = blue
   }
\def\be{\begin{equation}}
\def\ee{\end{equation}}
\def\bea{\begin{eqnarray}}
\def\eea{\end{eqnarray}}

\def\be{\begin{equation}}
\def\ee{\end{equation}}
\def\bea{\begin{eqnarray}}
\def\eea{\end{eqnarray}}

\begin{document}

\title{Exact black hole solutions with semi-symmetric vectorial torsion}
\author{Lehel Csillag}
\affiliation{Department of Physics, Babeș-Bolyai University, Kogălniceanu
Street, Cluj-Napoca 400084, Romania} 
\affiliation{Faculty of Mathematics and
Computer Science, Transilvania University, Iuliu Maniu Street 50, Brașov
500091, Romania}
\email{lehel.csillag@ubbcluj.ro,lehel.csillag@unitbv.ro}
\author{Tiberiu Harko}
\email{tiberiu.harko@aira.astro.ro}
\affiliation{Department of Physics, Babeș-Bolyai University, Kogălniceanu
Street, Cluj-Napoca 400084, Romania} 
\affiliation{Astronomical Observatory,
19 Cireșilor Street, Cluj-Napoca 400487, Romania}

\begin{abstract}
Semi-Symmetric metric gravity  extends general relativity by
introducing a semi-symmetric connection with vectorial torsion. The modified
field equations contain extra contributions from the torsion vector and its
derivatives. Assuming a purely radial torsion vector consistent with
spherical symmetry, we derive two exact vacuum solutions. The metric depends
on two integration constants, defining the Schwarzschild radius and a new
cosmological-type length scale. {Classical Solar System tests, including perihelion precession, light deflection, and radar echo delay, are used to constrain the new length scale and the associated deviations from general relativity. The thermodynamic properties of the resulting black hole solutions are investigated, and it is shown that the vector perturbation sector is linearly stable.}
\end{abstract}

\date{\today }
\maketitle

\affiliation{Department of Physics, Babeș-Bolyai University, Kogălniceanu
Street, Cluj-Napoca 400084, Romania} 
\affiliation{Astronomical Observatory,
19 Cireșilor Street, Cluj-Napoca 400487, Romania}

{\ \hypersetup{linkcolor=blue} }

\section{Introduction}

\hypersetup{citecolor=blue}

Black-hole solutions of gravitational field theories represent one of the
main tools for testing theoretical predictions against observations. The
first exact solution of general relativity was the spherically symmetric
vacuum solution of the Einstein field equations, obtained over a century ago 
\cite{1}. The Schwarzschild solution describes the gravitational field of a
non-rotating compact object and introduced the concept of black holes as a
new class of stellar-like objects in astrophysics. The exterior vacuum
solution for a rotating compact object was obtained by Kerr almost fifty
years after the Schwarzschild metric \cite{Kerr}. Investigations of the
Schwarzschild and Kerr solutions have raised fundamental theoretical
questions, regarding black hole singularities, which remain open \cite{1a,1b}%
.

One of the most important results in relativistic astrophysics is that
compact objects with masses greater than 3-4 solar masses must be black
holes \cite{2}. However, the first observational evidence for the existence
of black holes came relatively recently, with the observational proof that
the High Mass Binary Object Cyg X-1 may be a black hole \cite{3}. The
discovery of super-massive black holes, which are present at the center of
each galaxy \cite{4}, opened some new observational perspectives on the
nature and observational appearance of these objects \cite{5}. The first
image of the supermassive black hole M87* by the Event Horizon Telescope
collaboration \cite{6,7,8} marked a significant advance in the study of
ultra-compact massive objects. Observations of Sgr A* are consistent with a
Kerr-like geometry, but they do not yet definitively rule out possible
deviations from general relativity.

Many black hole solutions have been obtained in general relativity and in
modified gravity theories (for a review, see \cite{9}). Black holes are
generally understood to form as the end state of a gravitational collapse of
initially regular matter distributions. This raises the question of which
initial conditions lead to black hole formation and which may produce
alternative outcomes, such as naked singularities.

The possible existence of a  naked singularity would provide counterexamples to
the cosmic censorship hypothesis (CCH), proposed by Penrose \cite{10}, which
essentially states that curvature singularities in \textit{asymptotically
flat space-times} are always hidden behind event horizons. This happens,
according to Penrose, due to the existence of a cosmic censor, which outlaws
the birth of naked singularities, hiding all of them in an absolute event
horizon. The conjecture can be formulated either in a strong sense (in a
reasonable geometry naked singularity simply cannot exist), or in a weak
sense (even if such singularities do exist, they are covered by an event
horizon). For a detailed discussion of CCH see \cite{11}.

The rapid development of modified gravity theories \cite{12,13,14,15}, many of which are based on geometrical structures beyond the Riemannian framework of general relativity, has stimulated considerable interest in vacuum solutions of the corresponding field equations. Particular attention has been devoted to black hole solutions in Poincar\'e gauge theory (PGT), in which curvature and torsion may both contribute dynamically to the gravitational field \cite{16}. Exact vacuum solutions with nonvanishing torsion in PGT have been obtained in \cite{17,18,19,20,21}. Black hole solutions have also been investigated in $f(T)$ gravity, where the gravitational interaction is formulated in terms of torsion rather than curvature \cite{22,23,24}. Beyond compact-object physics, torsion has been considered in cosmological applications, including alternative inflationary scenarios \cite{McInnes2025} and possible geometrical explanations of the observed anisotropy of the Hubble expansion \cite{McInnes2025NPB}. Related extensions involving nonmetricity have also been explored.  Exact and numerical black hole solutions in Weyl geometric gravity have been reported in \cite{25,26}.

In 1924, Friedmann and Schouten introduced a class of affine connections now known as semi-symmetric connections \cite{27}. Their mathematical properties have been studied in a variety of geometrical settings \cite{28,29,30,Maksimovic2025,Maksimovic2025PerfectFluid}, and semi-symmetric connections have also appeared in several physical applications, although sometimes with different interpretations \cite{31,32,Chaudhary2024}. In semi-symmetric metric geometry, the torsion tensor is completely determined by a single covector field, or dually by a vector field, which is usually referred to as the torsion vector. A systematic analysis of the gravitational and cosmological implications of semi-symmetric metric connections was presented in \cite{32}, where the field equations of semi-symmetric metric gravity were formulated and several cosmological solutions were obtained.

In this work, we derive the exact vacuum solutions of the field equations of
semi-symmetric metric gravity, as formulated in \cite{32}, under the
assumption of static spherical symmetry. The field equations admit two exact
solutions, which generalize the Schwarzschild solution of general relativity
and are characterized by a single new length scale, $R_g$, and its sign. To
test the viability of these solutions, we examine three classical Solar
System probes: the perihelion precession of Mercury, the deflection of light
by the Sun, and the radar time delay. These tests provide lower bounds on
the free parameter of the solutions.

The paper is organized as follows. In Section~\ref{sect1}, we review the
fundamentals of semi-symmetric metric gravity and present the field
equations under spherical symmetry. The exact vacuum solutions of these
equations are derived in Section~\ref{sect2}. In Section~\ref{sect3}, we
examine Solar System tests in this metric and derive constraints on the
model parameter $R_g$. {The thermodynamic and perturbative properties of the solutions are examined in Section \ref{sec:properties}.} Finally, we summarize and discuss our results in
Section~\ref{sect4}.

\section{Semi-symmetric metric gravity: geometry, and field equations}

\hypersetup{citecolor=blue}\label{sect1}
{In this section, we briefly review semi-symmetric metric gravity. Its torsion tensor is restricted to the vectorial form \eqref{semisymmetric}, and is therefore completely determined by a single covector field with components $\pi_\mu$. This reduces the $24$ independent components of a general torsion tensor in four dimensions to $4$ components, making the theory more constrained and arguably the simplest torsional extension of general relativity.}

{Semi-symmetric metric gravity differs from  Einstein--Cartan theory \cite{EC}, where torsion is algebraically sourced by the spin density of matter and generally vanishes in vacuum. It is also more restricted than Poincaré gauge gravity, which may contain several irreducible dynamical torsion modes together with multiple coupling constants \cite{16}. As noted above, in semi-symmetric metric gravity only the vectorial part of the irreducible decomposition enters. This is in sharp contrast to teleparallel and $f(T)$ theories, where curvature vanishes and gravity is described entirely through torsion (also in vacuum, this is a similarity) \cite{22,23,24}.  Semi-symmetric metric gravity is metric compatible, and generally possesses both nonzero curvature and torsion. The modifications of the Einstein gravitational field equations arise from the restricted form of the connection, rather than from adopting an arbitrary function of a torsion invariant. }

{Semi-symmetric metric gravity is undoubtedly an interesting theory among torsional theories, whose validation can be provided only through extensive comparisons with observational and even experimental data.  Its main advantage is its simplicity: the torsion being completely determined by a vector field. The deviations from the Schwarzschild solution are simply governed by one additional length scale, $R_g$. In cosmology, the same torsion sector modifies the Friedmann equations, and may act as an effective source of dark energy \cite{31,32,Chaudhary2024}, thus providing a geometric explanation of the late acceleration of the Universe, and of the observational data. 
}

%\subsection{The semi-symmetric metric connection}

\paragraph{Semi-symmetric metric geometry.}

Friedmann and Schouten \cite{27} defined a semi-symmetric connection as an
affine connection whose torsion tensor \footnote{%
We adopt the convention in which the torsion of an affine connection is
defined as $\tensor{T}{^\mu_\nu_\rho} := 2 
\tensor{\Gamma}{^{\mu}_{[\rho
\nu]}}. $} satisfies 
\begin{equation}  \label{semisymmetric}
\tensor{T}{^\mu_\nu _\rho} = \pi_{\rho} \delta^{\mu}_{\nu} - \pi_{\nu}
\delta^{\mu}_{\rho} ,
\end{equation}
where $\pi_\rho$ are the components of an arbitrary one-form. When the
nonmetricity vanishes, the connection is referred to as a semi-symmetric
metric connection.

 A torsionful {metric-compatible} affine connection can be written as 
\begin{equation}  \label{generalconnection}
\tensor{{\Gamma}}{^\mu _\nu _\rho}=\tensor{\gamma}{^\mu _\nu _\rho} - \frac{1%
}{2}g^{\lambda \mu}(T_{\rho \nu \lambda}+T_{\nu \rho \lambda}- T_{\lambda
\rho \nu}) ,
\end{equation}
where $\tensor{\gamma}{^\lambda _\mu _\nu}$ denotes the Christoffel symbols
of the Levi-Civita connection. By assuming the torsion takes the
semi-symmetric form \eqref{semisymmetric}, the connection coefficient
functions of semi-symmetric metric geometry become \cite{32} 
\begin{equation}  \label{Christoffelsemisymmetric}
\tensor{{\Gamma}}{^\mu _\nu _\rho}= \tensor{\gamma}{^\mu _\nu _\rho} -
\pi^{\mu} g_{\rho \nu} + \pi_{\nu} \delta^{\mu}_{\rho} .
\end{equation}
The Riemann tensor associated with this connection is given by \cite{32} 
\begin{eqnarray}  \label{riemanncurvaturesemisym}
\tensor{{R}}{^\mu_{\nu \rho \sigma}}&=&\overset{\circ}{{R}} \tensor{}{^\mu
_\nu _\rho _\sigma}- S_{\sigma \nu} \delta^{\mu}_{\rho}+S_{\rho \nu}
\delta^{\mu}_{\sigma}  \notag \\
&-& g_{\sigma \nu} S_{\rho \lambda} g^{\lambda \mu}+g_{\rho \nu} S_{\sigma
\lambda} g^{\lambda \mu} ,
\end{eqnarray}
where  $\overset{\circ}{{R}}\tensor{}{^\mu _\nu _\rho _\sigma}$ is the
Riemann tensor of the Levi-Civita connection, and 
\begin{equation}  \label{stensorpitensor}
S_{\nu \sigma}=\overset{\circ}{\nabla}_{\nu} \pi_{\sigma}- \pi_{\nu}
\pi_{\sigma} + \frac{1}{2} g_{\nu \sigma} \pi_{\lambda} \pi^{\lambda} .
\end{equation}
The Ricci tensor is readily obtained as 
\begin{equation}  \label{Riccicurvaturesemisym}
R_{\nu \sigma}=\overset{\circ}{R}_{\nu \sigma} -2 S_{\sigma \nu} -g_{\nu
\sigma} S_{\lambda \beta} g^{\lambda \beta} ,
\end{equation}
while the Ricci scalar reads 
\begin{equation}  \label{Ricciscalarsemisym}
R=\overset{\circ}{R} - 6 S_{\beta \lambda} g^{\beta \lambda}.
\end{equation}

\paragraph{Field equations of semi-symmetric metric gravity.}

The generalized vacuum Einstein equations are \cite{32}
\begin{equation}
R_{(\nu \sigma)} - \frac{1}{2} R g_{\nu \sigma}=0 .
\end{equation}
Substituting Eq. \eqref{Riccicurvaturesemisym} and Eq. %
\eqref{Ricciscalarsemisym} leads to 
\begin{equation}  \label{einstein}
\overset{\circ}{R}_{\nu \sigma}- \frac{1}{2} g_{\nu \sigma} \overset{\circ}{R%
}-S_{\sigma \nu} - S_{\nu \sigma} +2 g_{\sigma \nu} S_{\lambda \beta}
g^{\lambda \beta}=0 .
\end{equation}
In the limit of $S$ tending to zero, the usual vacuum Einstein equations are
recovered.

Expressed in terms of the one-form $\pi$, {the field equations can be equivalently represented as}
\begin{eqnarray}  \label{Einsteinsemisymmetricequation}
\overset{\circ}{R}_{\nu \sigma}-\frac{1}{2} g_{\nu \sigma} \overset{\circ}{R}
&-& \overset{\circ}{\nabla}_{\sigma} \pi_{\nu} - \overset{\circ}{\nabla}%
_{\nu} \pi_{\sigma}+2 \pi_{\sigma} \pi_{\nu}+  \notag \\
&+& 2 g_{\sigma \nu} \overset{\circ}{\nabla}_{\lambda} \pi^{\lambda} +
g_{\nu \sigma} \pi^{\rho}\pi_{\rho}=0 .
\end{eqnarray}

\section{Exact black hole solutions in semi-symmetric metric gravity}

\label{sect2}

In this section, we derive an exact vacuum solution of the gravitational
field equations for a spherically symmetric configuration in semi-symmetric
gravity, generalizing the Schwarzschild solution of general relativity.

\paragraph{Field equations in spherical symmetry.}

For the spacetime metric, we assume the standard form 
\begin{equation}
ds^{2}=e^{\nu (r)}c^{2}dt^{2}-e^{\lambda (r)}dr^{2}-r^{2}\left( d\theta
^{2}+\sin ^{2}\theta d\phi ^{2}\right),
\end{equation}
where the metric functions $\nu$ and $\lambda$ depend only on the radial
coordinate $r$. For the torsion vector, we assume a purely radial component 
\begin{equation}
\pi ^{\lambda }=\left( 0,\pi ^{1}(r),0,0\right),
\end{equation}
so that\footnote{%
From now on, we do not write out the $r$-dependence of $\pi^{1}$ to avoid
cumbersome notation, but it should always be understood that $%
\pi^{1}=\pi^{1}(r).$} 
\begin{equation}
\pi _{\rho }\pi ^{\rho }=g_{\alpha \beta }\pi ^{\alpha }\pi ^{\beta
}=-e^{\lambda (r)}\pi ^{1}\pi ^{1}.
\end{equation}

With these assumptions we obtain a set of three differential equations for
the metric tensor components, and for the torsion vector 
\begin{equation}
-e^{-\lambda }\left( \frac{1}{r^{2}}-\frac{\lambda ^{\prime }}{r}\right) +%
\frac{1}{r^{2}}+2\frac{\partial \pi ^{1}}{\partial r}+\frac{4\pi ^{1}}{r}%
+\lambda ^{\prime }\pi ^{1}-e^{\lambda}\pi ^{1}\pi ^{1}=0,  \label{1}
\end{equation}
\begin{equation}
-e^{-\lambda }\left( \frac{\nu ^{\prime }}{r}+\frac{1}{r^{2}}\right) +\frac{1%
}{r^{2}}+\frac{4\pi ^{1}}{r}+\nu ^{\prime }\pi ^{1}-3e^{\lambda }\pi ^{1}\pi
^{1}=0,  \label{2}
\end{equation}
\begin{eqnarray}
&-&\frac{1}{2}e^{-\lambda }\left( \nu ^{\prime \prime }+\frac{\nu ^{\prime 2}%
}{2}+\frac{\nu ^{\prime }-\lambda ^{\prime }}{r}-\frac{\nu ^{\prime }\lambda
^{\prime }}{2}\right) +2\frac{\partial \pi ^{1}}{\partial r}+\frac{2\pi ^{1}%
}{r}  \notag \\
&&+\left( \nu ^{\prime }+\lambda ^{\prime }\right) \pi ^{1}-e^{\lambda }\pi
^{1}\pi ^{1}=0.  \label{3}
\end{eqnarray}
In the following, for simplicity, we denote $\pi ^{1}=\psi $.

\subsection{Solving the gravitational field equations}

By subtracting Eqs. (\ref{1}) and (\ref{2}) we obtain the relation 
\begin{equation}
e^{-\lambda }\frac{\nu ^{\prime }+\lambda ^{\prime }}{r}+2 \psi ^{\prime }+
\left( \lambda ^{\prime }-\nu ^{\prime }\right) \psi +2 e^{\lambda }\psi
^{2}=0.  \label{5}
\end{equation}

We look for solutions that satisfy the condition $\nu +\lambda =0$,
similarly to the Schwarzschild solution and other vacuum configurations.
Then Eq. (\ref{5}) simplifies to 
\begin{equation}
\psi ^{\prime }+\lambda ^{\prime }\psi + e^{\lambda }\psi ^{2}=0,
\end{equation}%
or equivalently, 
\begin{equation}
\lambda ^{\prime }+ e^{\lambda }\psi +\frac{\psi ^{\prime }}{\psi }=0.
\label{6}
\end{equation}

Introducing a new variable $f$ via $\lambda =\ln f$ so that $\lambda
^{\prime }=f^{\prime }/f$, Eq.~(\ref{6}) becomes 
\begin{equation}
\frac{f^{\prime }}{f^{2}}+\frac{\psi ^{\prime }}{\psi }\frac{1}{f}+ \psi =0.
\label{7}
\end{equation}
Defining $u=1/f$ with $u'=-\frac{f'}{f^2}.$, Eq.~(\ref{7}) transforms
into 
\begin{equation}
u^{\prime }=\frac{\psi ^{\prime }}{\psi }u+ \psi ,  \label{8}
\end{equation}%
which has the general solution 
\begin{equation}  \label{9}
u=\left( C+ r\right) \psi ,
\end{equation}%
where $C$ is an arbitrary integration constant. Hence, 
\begin{equation}
e^{\lambda }=f=\frac{1}{\left( C+ r\right) \psi },e^{-\lambda }=\left( C+
r\right) \psi .
\end{equation}

Substituting these expressions for $e^{-\lambda}$ and $\lambda^{\prime }$
into Eq.(\ref{1}), we obtain for $\psi$ the differential equation 
\begin{equation}
\psi ^{\prime }+\frac{C-r}{r(C+r)}\psi -\frac{1}{Cr}=0,
\end{equation}%
with the general solution 
\begin{equation}
\psi =\frac{C_{1}(C+r)^{2}}{r}-\frac{C+r}{Cr},
\end{equation}%
where $C_{1}$ is an integration constant. Hence, the metric functions are given
by 
\begin{equation}
e^{\lambda }=\frac{Cr}{(C+r)^{2}\left[ C_{1}C(C+r)-1\right] },
\end{equation}
and 
\begin{equation}  \label{33}
e^{\nu }=C_{1}r^{2}\left( 1+\frac{C}{r}\right) ^{2}\left[ 1+%
\frac{1}{C_{1}}\left( C_{1}C-\frac{1}{C}\right) \frac{1}{r}\right] .
\end{equation}
Eqs.~(\ref{33}) provide a bi-parametric solution of the vacuum spherically
symmetric field equations in semi-symmetric metric gravity. Direct
substitution of the metric functions and $\psi$ into Eqs. ~(\ref{1})--(\ref%
{3}) verifies that all three equations are satisfied. The integration
constant $C$ has dimensions of length and is conventionally written as $%
C=\varepsilon R_{g}$ with $\varepsilon=\pm 1, R_{g} >0$. By analogy with the
Schwarzschild metric we define 
\begin{equation}
\frac{1}{C_{1}}\left( C_{1}C-\frac{1}{C}\right) =\frac{C_{1}C^{2}-1}{CC_{1}}
\equiv-r_{g},  \label{10}
\end{equation}
where $r_{g}=2GM/c^{2}$ is the gravitational radius of the object. Then $%
C_{1}$ must have the dimension of $1/$length$^{2}$, giving

\begin{equation}
C_{1}=\frac{1}{R_{g}^{2}\left( 1 + \varepsilon \frac{r_{g}}{R_{g}}\right) }.
\end{equation}

Hence, the metric takes the final form 
\begin{equation}  \label{fin}
e^{\nu }=\frac{1}{\left( 1 + \varepsilon \frac{r_{g}}{R_{g}}%
\right) }\left( \frac{r}{R_{g}}\right) ^{2}\left( 1 + \varepsilon \frac{R_{g}%
}{r}\right) ^{2}\left( 1-\frac{r_{g}}{r}\right) .
\end{equation}

The solution depends on two length scales: the usual gravitational radius $%
r_{g}$, and a new scale $R_g$. In the limits $R_g/r \gg 1$, or $R_g \gg r$,
and $r_g/R_g \ll 1$, the Schwarzschild metric of general relativity is
recovered. Explicitly, the metric (\ref{fin}) can also be written as 
\begin{equation}  \label{37}
e^{\nu}=\frac{1}{1 + \varepsilon \frac{r_g}{R_g}} \left[1-2
\varepsilon \frac{r_g}{R_g}-\frac{r_g}{r}-\frac{r_g - 2 \varepsilon R_g}{%
R_g^2}r+\frac{r^2}{R_g^2}\right].
\end{equation}

The torsion vector is 
\begin{equation}  \label{38}
\psi (r) =\frac{1}{R_{g}}\left( 1 + \varepsilon \frac{R_{g}}{r}\right) \frac{%
r-r_{g}}{R_{g} +\varepsilon r_{g}}.
\end{equation}

Finally, Eq.~(\ref{37}) can be rewritten in a Schwarzschild-like form 
\begin{equation}
e^{-\lambda}=e^{\nu}=\frac{1}{1+ \varepsilon \frac{r_g}{R_g}}\left(1-\frac{%
r_{g}+r_g^{(eff)}(r)}{r}\right),
\end{equation}
where the effective gravitational radius is defined as 
\begin{equation}
\begin{aligned} r_g^{(eff)}(r)&=\frac{2GM_{eff}(r)}{c^2}\\
&=r_gr\left(\frac{2 \varepsilon}{R_g}+\frac{r_g -2 \varepsilon
R_g}{r_gR_g^2}r-\frac{r^2}{r_gR_g^2}\right), \end{aligned}
\end{equation}
with $M_{eff}(r)$ denoting the effective mass associated with torsional
effects. The effective geometric density is 
\begin{equation}
\begin{aligned} \rho_{eff}(r)&=\frac{1}{4 \pi r^2} \frac{d M_{eff}(r)}{dr}\\
&=\frac{c^2}{8 \pi G} r_{g} \left[\frac{2 \varepsilon}{R_{g}
r^2}+\frac{2\left(r_g-2 \varepsilon R_{g} \right)}{r_{g} R_{g}^2}
\frac{1}{r} - \frac{3}{r_g R_g^2} \right]. \end{aligned}
\end{equation}

We introduce now the notation $\delta =r_{g}/R_{g}$, which allows to rewrite
the metric (\ref{37}) in the form 
\begin{equation}
e^{-\lambda }=e^{\nu }=\frac{1}{1+\varepsilon \delta }\left[ 1-2\varepsilon
\delta -\frac{r_{g}}{r}-\delta \left( \delta -2\varepsilon \right) \frac{r}{%
r_{g}}+\delta ^{2}\frac{r^{2}}{r_{g}^{2}}\right] .  \label{fin1}
\end{equation}

\section{The Solar System tests}

\label{sect3}

The three standard Solar System tests of gravity are the perihelion
precession of Mercury, the deflection of light by the Sun, and the radar
echo delay. These tests have been used to verify the Schwarzschild solution.
We apply them here to the exact solution~(\ref{fin}) of semi-symmetric
metric gravity.

\subsection{The perihelion advance}

We first consider constraints on the parameter $R_g$ of the exact
semi-symmetric black hole solution derived from the perihelion precession of
a planet.

\paragraph{The perihelion advance of a planet: general formalism.}

In the metric given by Eq. (\ref{fin}), the motion of a test particle is
obtained from the variational principle 
\begin{equation}
\delta \int \sqrt{e^{\nu }c^{2}\dot{t}^{2}-e^{\lambda }\dot{r}
^{2}-r^{2}\left( \dot{\theta}^{2}+\sin ^{2}\theta \dot{\phi}^{2}\right) }
ds=0,  \label{var}
\end{equation}
where the dot denotes $d/ds$. Since the orbit is planar, we take $\theta
=\pi /2$, and we use $\phi $ as the angular coordinate. Neither $t$ nor $%
\phi $ do explicitly appear in Eq. (\ref{var}), and hence we obtain two
constants of motion, $e^{\nu }c^{2}\dot{t}=E=\mathrm{constant}$, and $r^{2}%
\dot{\phi}=L=\mathrm{constant}$. $E$ is the energy of the particle, while $L 
$ is the angular momentum.

From the line element Eq. (\ref{fin}), the radial equation of motion is 
\begin{equation}
\dot{r}^{2}+e^{-\lambda }r^{2}\dot{\phi}^{2}=e^{-\lambda }\left( e^{\nu
}c^{2}\dot{t}^{2}-1\right).  \label{constants}
\end{equation}

Using the conserved energy $E$ and angular momentum $L$, this becomes 
\begin{equation}
\dot{r}^{2}+e^{-\lambda }\frac{L^{2}}{r^{2}}=e^{-\lambda }\left(\frac{E^{2} 
}{c^{2}}e^{-\nu }-1\right) .  \label{inter1}
\end{equation}

Introducing $u = 1/r$ and $d/ds = L u^2 d/d\phi$, Eq.~(\ref{inter1}) reads 
\begin{equation}
\left( \frac{du}{d\phi }\right) ^{2}+e^{-\lambda }u^{2}=\frac{1}{L^{2}}
e^{-\lambda }\left( \frac{E^{2}}{c^{2}}e^{-\nu }-1\right) .
\end{equation}

By representing $e^{-\lambda }$ as 
\begin{equation}
e^{-\lambda }=\gamma \left[ 1-f(u)\right] ,  \label{repr}
\end{equation}%
where $\gamma $ is a constant, and taking into account that $\nu +\lambda =0$%
, we arrive at the equation 
\begin{equation}
\left( \frac{du}{d\phi }\right) ^{2}+\gamma u^{2}=\gamma f(u)\left( u^{2}+%
\frac{1}{L^{2}}\right) +\frac{1}{L^{2}}\left( \frac{E^{2}}{c^{2}C}-\gamma
\right) \equiv G(u),  \label{ueq_basic}
\end{equation}%
which, after taking its derivative with respect to $\phi $ becomes 
\begin{equation}
\frac{d^{2}u}{d\phi ^{2}}+\gamma u=F(u),  \label{inter2}
\end{equation}%
where $F(u)=\frac{1}{2}\frac{dG(u)}{du}$.

The radius of a circular orbit $u=u_{circ}$ is obtained as a root of the
equation $\gamma u_{circ}=F\left( u_{circ}\right) $. The deviations $\Delta
=u-u_{circ}$ from a circular orbit are obtained as solutions of the equation 
\begin{equation}
\frac{d^{2}\Delta }{d\phi ^{2}}+\left[ \gamma -\left( \frac{dF}{du}\right)
_{u=u_{circ}}\right] \Delta =O\left( \Delta ^{2}\right) ,
\end{equation}
obtained by substituting $u=u_{circ}+\Delta $ in Eq. (\ref{inter2}). Hence,
in the first order of approximation in $\Delta $, the equation of the
trajectory is given by 
\begin{equation}
\Delta =\Delta _{0}\cos \left( \sqrt{\gamma -\left( \frac{dF}{du}%
\right)_{u=u_{circ}} }\phi +\alpha \right) ,
\end{equation}
where $\Delta _{0}$ and $\alpha $ are constants of integration. The orbital
angle between successive perihelia is 
\begin{equation}
\phi =\frac{2\pi }{\sqrt{\gamma-\left( \frac{dF}{du}\right) _{u=u_{circ}}}}=%
\frac{ 2\pi }{1-\sigma },  \label{prec}
\end{equation}
where $\sigma $ is the perihelion advance. Expanding to first order, 
\begin{equation}
\sigma =1-\sqrt{\gamma-\left( \frac{dF}{du}\right) _{u=u_{circ}}}\approx 1-%
\sqrt{\gamma}+\frac{1}{2\sqrt{\gamma} }\left( \frac{dF}{du}\right)
_{u=u_{circ}}.
\end{equation}

Thus, for a complete orbital rotation, $\phi \approx 2\pi (1+\sigma)$,
giving a perihelion shift 
\begin{equation}
\Delta \phi\approx 2\pi \left(1-\sqrt{\gamma}\right)+\frac{\pi}{\sqrt{\gamma}%
}\left(\frac{dF(u)}{du}\right)_{u=u_{circ}},
\end{equation}
where we used $\Delta \phi= \phi - 2\pi \approx 2\pi \sigma$.

\paragraph{The Newtonian approximation.}

Assume the planet follows a Keplerian ellipse with semi-axes $a$ and $b = a 
\sqrt{1-e^2}$, where $e$ is the eccentricity. The oriented areal element is $%
d\vec{\sigma} = (\vec{r} \times d\vec{r})/2$, giving an areal velocity 
\begin{equation}
\left| \frac{d\vec{\sigma}}{dt} \right| = \frac{|\vec{r} \times d\vec{r}|}{2}
= \frac{r^2}{2} \frac{d\phi}{dt} \approx \frac{\pi a^2 \sqrt{1-e^2}}{T},
\end{equation}
where $T$ is the orbital period, $T^2 = 4\pi^2 a^3 / GM$.

For small velocities, $ds \approx c\, dt$, and the conserved relativistic
angular momentum gives $r^2 d\phi / dt = c L$. Hence, 
\begin{equation}
L = \frac{2 \pi a^2 \sqrt{1-e^2}}{c T}, \qquad \frac{1}{L^2} = \frac{c^2}{G
M a (1-e^2)}.
\end{equation}

\paragraph{Perihelion advance of Mercury.}

To analyze the perihelion precession of a planet in the black-hole solution of semi-symmetric metric gravity, we start from the line element given in Eq.~(\ref{fin1}). Matching this metric to the representation in Eq.~(\ref{repr}) yields  
\begin{equation}
\gamma = \frac{1}{1+\varepsilon\,\delta}.
\end{equation}
Assuming $\delta = r_g / R_g \ll 1$, we can approximate $\gamma \approx 1$. The conditions $|2\varepsilon\delta \ll| 1$ and $\delta \ll 1$ are also well satisfied within Solar--System scales. Keeping only first order terms in $\delta$, the metric at the level of the Solar System takes the approximate form
\begin{equation}
e^\nu = e^{-\lambda} \approx 1 - \frac{r_g}{r} + 2\varepsilon\delta\,\frac{r}{r_g}
= 1 - f(u).
\label{58}
\end{equation}

In this limit, the relevant functions become 
\begin{align}
f(u)& =r_{g}u-\frac{2\varepsilon \delta }{r_{g}}\frac{1}{u}, \\
G(u)& =\left( r_{g}u-\frac{2\varepsilon \delta }{r_{g}}\frac{1}{u}\right)
\left( u^{2}+\frac{1}{L^{2}}\right) +\frac{1}{L^{2}}\left( \frac{E^{2}}{c^{2}%
}- {1} \right) , \\
F(u)& =\frac{1}{2}\frac{dG(u)}{du}=\frac{3r_{g}u^{2}}{2}+\frac{r_{g}}{2L^{2}}%
-\frac{\delta \varepsilon }{r_{g}}+\frac{\delta \varepsilon }{L^{2}r_{g}}%
\frac{1}{u^{2}},
\end{align}

\begin{equation}
\pi \frac{dF(u)}{du}=3\pi r_{g}u-\frac{2\varepsilon \delta \pi }{L^{2}r_{g}}%
\frac{1}{u^{3}}.  \label{pp}
\end{equation}

The circular orbit satisfies the equation $u=F(u)$, which gives for the
orbit radius the algebraic equation 
\begin{equation}
\frac{3}{2}r_{g}u^{2}-u+\frac{r_{g}}{2L^{2}}-\frac{\delta \varepsilon }{r_{g}%
}+\frac{\delta \varepsilon }{L^{2}r_{g}}\frac{1}{u^{2}}=0.  \label{52}
\end{equation}

We neglect again the constant term $-\delta \varepsilon /r_{g}$ in the above
equation. Moreover, we
assume that the last term in Eq (\ref{52}), $P(u)=$ $\left( \varepsilon
\delta /L^{2}r_{g}\right) \left( 1/u^{2}\right) $, is a small perturbative
term, an approximation which is certainly valid for the Solar System.

The latter assumption allows us to solve Eq.~(\ref{52}) perturbatively.
Neglecting $P(u)$ yields the zeroth-order equation
\begin{equation}
\frac{3}{2} r_{g} u_{0}^{2} - u_{0} + \frac{r_{g}}{2L^{2}} = 0,
\end{equation}
whose physical root is
\begin{equation}
u_{0} = \frac{1 - \sqrt{1 - 3 r_{g}^{2}/L^{2}}}{3 r_{g}}
    \approx \frac{r_{g}}{2L^{2}},
\end{equation}
where the approximation $3 r_{g}^{2} / L^{2} \ll 1$ has been used.

Substituting $u_{0}$ into $P(u)$ yields the first-order correction 
\begin{equation}
\frac{3}{2}r_{g}u_{1}^{2}-u_{1}+\frac{r_{g}}{2L^{2}}+P(u_{0})=0,
\end{equation}%
with solution 
\begin{equation}
u_{1}\approx \frac{r_{g}}{2L^{2}}+P(u_{0})=\frac{r_{g}}{2L^{2}}+4\varepsilon
\delta \frac{L^{2}}{r_{g}^{3}}.
\end{equation}

The perihelion precession in semi-symmetric metric gravity is then obtained
from Eq. (\ref{pp}) as 
\begin{align}
\Delta \phi & =3\pi r_{g}u_{1}-\left. \frac{2\varepsilon \delta \pi }{%
L^{2}r_{g}}\frac{1}{u^{3}}\right\vert _{u\approx u_{0}}  \notag \\
& \approx \frac{3\pi r_{g}^{2}}{2L^{2}}+12\pi \varepsilon \delta \frac{L^{2}%
}{r_{g}^{2}}-16\pi \varepsilon \delta \frac{L^{4}}{r_{g}^{4}},
\end{align}%
where we have approximated $u_{1}\approx u_{0}$ in the second term.

Expressed in orbital parameters, this reads 
\begin{eqnarray}
\Delta \phi & =&\frac{6 \pi GM}{ac^{2}(1-e^{2})}+3\pi \varepsilon \delta 
\frac{c^{2}a\left( 1-e^{2}\right) }{GM}  \notag \\
&&-\pi \varepsilon \delta \frac{a^{2}\left( 1-e^{2}\right) ^{2}c^{4}}{%
G^{2}M^{2}} =\Delta \phi _{\mathrm{GR}}+\Delta \phi _{\mathrm{cor}},
\end{eqnarray}
where $\Delta \phi _{\mathrm{GR}}$ is the standard general relativistic
contribution and $\Delta \phi _{\mathrm{cor}}$ is the correction due to
torsion.

The numerical value of the perihelion precession of the planet Mercury,
obtained from observations, is $\Delta \phi _{Obs}=43.11\pm 0.21$ arcsec per
century \cite{Pa1,Pa2}. From the general relativistic formula for the
perihelion precession of Mercury, we obtain $\Delta \phi
_{GR}=42.94$ arcsec per century. Hence, the difference 
\begin{equation}
\Delta \phi _{diff}=\Delta \phi _{Obs}-\Delta \phi _{GR}=0.17\pm 0.21\;%
\mathrm{arcsec/century},
\end{equation}
can be attributed to other physical effects. We assume that the deviations
from general relativity are entirely due to the presence of torsion.
Thus, $\Delta \phi _{cor}\approx \Delta \phi _{diff}$, which allows us to
estimate the correction term as 
\begin{equation}
3\varepsilon \delta \pi \frac{c^{2}\left( 1-e^{2}\right) a}{GM}\left[ 1-%
\frac{c^{2}\left( 1-e^{2}\right) a}{3GM}\right] \approx \Delta \phi _{diff},
\end{equation}%
or, by taking into account the definition of $\delta $, we obtain

\begin{equation}
6 \pi \varepsilon \frac{\left( 1-e^{2}\right) a}{R_{g}}\left[ 1-\frac{%
c^{2}\left( 1-e^{2}\right) a}{3GM}\right] \approx \Delta \phi _{diff}.
\end{equation}

In the following we adopt for the orbit parameters the parameters of the
planet Mercury, $a=57.91\times 10^{11}$ cm, $e=0.205615$, $M_{\odot
}=1.989\times 10^{33}$ g, while for the physical constants we take the
numerical values $c=2.998\times 10^{10}$ cm/s, and $G=6.67\times 10^{-8}$ cm$%
^{3}$/g/s$^{2}$, we also take into account that $1\;\mathrm{arcsec}%
=4.8484\times 10^{-6}$ radian.

In the case of the planet Mercury the term $c^{2}\left( 1-e^{2}\right)
a/3GM=1.2525\times 10^{7}$ is much greater than one, and hence we obtain for
the free parameter of the semi-symmetric metric black hole the estimate 
\begin{equation}
R_{g}\approx - 2\varepsilon \frac{c^{2}a^{2}\left( 1-e^{2}\right) ^{2}}{%
GM\Delta \phi _{diff}}.
\end{equation}

From the above equation it follows that the sign and the value of $%
\varepsilon $ must be fixed in the metric as $\varepsilon =-1$. As for the
numerical value of $R_{g}$, the perihelion precession of the planet Mercury
gives for it the estimate 
\begin{equation}
R_{g}\approx 5.0572\times 10^{26}\; \mathrm{cm}.
\end{equation}

\subsection{The deflection of light}

We consider now photon motion in the metric (\ref{58}) along null geodesics, 
$ds^2 = 0$. Derivatives with respect to an arbitrary affine parameter are
denoted by a dot. As in the case of massive particles, the motion admits two
conserved quantities: the energy $E$ and the angular momentum $L$.

From $ds^2=0$, the radial equation of motion reads 
\begin{equation}
\dot{r}^{2}+e^{-\lambda }r^{2}\dot{\phi}^{2}=e^{\nu -\lambda }c^{2}\dot{t}
^{2},
\end{equation}
which, using the constants of motion, becomes 
\begin{equation}
\dot{r}^{2}+e^{-\lambda}\frac{L^{2}}{r^{2}}=\frac{E^{2}}{c^{2}}e^{-\nu
-\lambda }.
\end{equation}

Introducing $r = 1/u$ and using $\dot{\phi}$ to eliminate the affine
derivative, with $e^{-\lambda} = \gamma (1-f(u))$, gives 
\begin{equation}
\left( \frac{du}{d\phi }\right) ^{2}+\gamma u^{2}=\gamma f(u)u^{2}+\frac{1}{%
c^{2}}\frac{ E^{2}}{L^{2}}e^{-\nu -\lambda }\equiv S(u)\, ,
\label{P_eq_basic}
\end{equation}
and differentiating with respect to $\phi$ yields 
\begin{equation}
\frac{d^2 u}{d\phi^2} + \gamma u = V(u), \quad \text{where} \quad V(u) = 
\frac{1}{2} \frac{dS}{du}.
\end{equation}

In the leading-order approximation, neglecting $V(u)$, the solution is 
\begin{equation}
u=\frac{\cos (\sqrt{\gamma }\,\phi )}{b},
\end{equation}%
representing a straight-line trajectory with impact parameter $b$.

At next order, substituting this solution into $V(u)$ gives the
inhomogeneous equation 
\begin{equation}
\frac{d^{2}u}{d\phi ^{2}}+u=V\left( \frac{\cos (\sqrt{\gamma }\,\phi )}{b}%
\right) ,
\end{equation}%
whose solution $u(\phi )$ determines the light trajectory. The asymptotic
angles are $\phi =-(\pi /2+\xi )$ on approach and $\phi =\pi /2+\xi $ on
exit, with the deflection angle $\Delta \phi _{LD}=2\xi $, found from the
condition $u(\pi /2+\xi )=0$, where we assume $\xi \ll 1$.

In the following we again adopt the approximation $\gamma \approx 1$. As in the previous case, the function $f(u)$ becomes
\begin{equation}
f(u)=r_{g}u-\frac{2\varepsilon \delta }{r_{g}}\frac{1}{u}.
\end{equation}%
Then, we immediately find 
\begin{equation}
S(u)=\left( r_{g}u-\frac{2\varepsilon \delta }{r_{g}}\frac{1}{u}\right)
u^{2}+\frac{E^{2}}{c^{2}L^{2}},
\end{equation}%
and 
\begin{equation}
V(u)=\frac{1}{2}\frac{dS(u)}{du}=\frac{3}{2}r_{g}u^{2}-\frac{\varepsilon
\delta }{r_{g}},
\end{equation}%
respectively. By substituting $u\rightarrow \cos (\phi )/b$ in the
expression of $V(u)$, we obtain for $u$ the linear inhomogeneous equation
\begin{equation}
\frac{d^{2}u}{d\phi ^{2}}+u-\frac{3r_{g}\cos ^{2}\phi }{2b^{2}}+\frac{\varepsilon \delta }{r_{g}}=0,
\end{equation}%
with the general solution 
\begin{equation}
u(\phi )=\frac{1}{b}\cos \phi -\frac{r_{g}}{4b^{2}}\cos (2\phi )+\frac{3r_{g}%
}{4b^{2}}-\frac{\varepsilon \delta }{r_{g}}.  \label{76}
\end{equation}%
The deflection angle $\Delta \phi _{LD}$ is obtained from the relation $%
u(\pi /2+\xi )=0$, giving in first order 
\begin{equation}
\Delta \phi _{LD}=\frac{2r_{g}}{b}\left[ 1- \varepsilon \delta \frac{b^{2}}{%
r_{g}^{2}}\right] =\Delta \phi _{LD}^{(GR)}\left( 1+\Delta \phi _{LD}^{(corr)} \right).
\label{77}
\end{equation}

The first term is the standard GR result, $\Delta \phi
_{LD}^{(GR)}=4GM/(bc^{2})$. Observational data from long-baseline radio
interferometry \cite{DL1,DL2} give $\Delta \phi _{LD}^{(GR)}=1.7510$ arcsec
and $|\Delta \phi _{LD}^{(corr)}|\leq 0.0002\pm 0.0008$. Hence, by assuming
that the entire difference between the general relativistic prediction and
observations is due to the presence of torsion, we obtain 
\begin{equation}
-\varepsilon\frac{b^2}{r_g R_g}=|\Delta \phi _{LD}^{(corr)}|.
\end{equation}

The above equation fixes again the sign of $\varepsilon $ to $-1$, and gives for the value of $R_{g}$ the estimate 
\begin{equation}
R_{g}\approx \frac{b^2}{r_g |\Delta \phi _{LD}^{(corr)}|}=\frac{b^2 c^2}{2 GM |\Delta \phi _{LD}^{(corr)}|}.
\end{equation}

Using $b=6.96340\times 10^{10}$ cm, $r_{g}=2.95376 \times 10^{5}$ cm, and $%
|\Delta \phi _{LD}^{(corr)}|=0.0002\times $ $4.8484\times 10^{-6}$ radian yields

\begin{equation}
R_{g} \approx 1.6929 \times 10^{25} \; \mathrm{cm}.
\end{equation}

\subsection{Radar echo delay}

The radar echo delay \cite{Pa1,Pa2} is the third important Solar System
test of general relativity. It measures the travel time of radar signals to
an inner planet or satellite in two situations: (a) when the signal passes
close to the Sun, and (b) when the signal travels far from the Sun.

\paragraph{The general formalism.}

Without the Sun's gravitational field, the light travel time between two
planets is 
\begin{equation}
T_0 = \int_{-l_1}^{l_2} \frac{dy}{c},
\end{equation}
where $l_1$ and $l_2$ are the distances of the planets from the Sun.

In the presence of the Sun's gravitational field, the travel time becomes 
\begin{equation}
T = \int_{-l_1}^{l_2} \frac{dy}{v} = \frac{1}{c} \int_{-l_1}^{l_2}
e^{[\lambda(r) - \nu(r)]/2} \, dy,
\end{equation}
where $v = c \, e^{(\nu-\lambda)/2}$ is the effective speed of light.

The corresponding time delay is 
\begin{equation}
\delta T = T - T_0 = \frac{1}{c} \int_{-l_1}^{l_2} \left\{ e^{[\lambda(r) -
\nu(r)]/2} - 1 \right\} dy.
\end{equation}

Using $r=\sqrt{y^{2}+R^{2}}$, we obtain 
\begin{equation}
\delta T=\frac{1}{c}\int_{-l_{1}}^{l_{2}}\left\{ e^{\frac{1}{2}\left[
\lambda \left( \sqrt{y^{2}+R^{2}}\right) -\nu \left( \sqrt{y^{2}+R^{2}}%
\right) \right] }-1\right\} dy.  \label{delay_eq}
\end{equation}

If the metric tensor components satisfy the condition $\nu +\lambda =0$, the
time delay is given by

\begin{equation*}
\delta T=\frac{1}{c}\int_{-l_{1}}^{l_{2}}\left\{ e^{-\nu \left( \sqrt{%
y^{2}+R^{2}}\right) }-1\right\} dy.
\end{equation*}

\paragraph{The Schwarzschild case.}

For the Schwarzschild metric, $\lambda = -\nu$, so that 
\begin{equation*}
e^{\lambda/2 - \nu/2} = e^{-\nu} = \left(1 - \frac{2GM}{c^2 r}\right)^{-1}
\approx 1 + \frac{2GM}{c^2 r}.
\end{equation*}
Hence, the radar time delay is 
\begin{align}
\delta T_{RD}^{(Sch)} &= \frac{2GM}{c^3} \int_{-l_1}^{l_2} \frac{dy}{\sqrt{%
y^2 + R^2}} = \frac{2GM}{c^3} \ln \frac{\sqrt{R^2 + l_2^2} + l_2}{\sqrt{R^2
+ l_1^2} - l_1}  \notag \\
&\approx \frac{2GM}{c^3} \ln \frac{4 l_1 l_2}{R^2},
\end{align}
where the approximation holds for $R^2/l_1^2 \ll 1$ and $R^2/l_2^2 \ll 1$.

For the Earth-Sun system, with $l_{1}=1.525\times 10^{13}$ cm (Earth-Sun
distance), $l_{2}=2.491\times 10^{13}$ cm (Mars-Sun distance), and $%
R=6.96340\times 10^{10}$ cm the Solar radius, we obtain 
\begin{equation}
\delta T_{RD}^{(Sch)}=\frac{2GM}{c^{3}}\ln \frac{4l_{1}l_{2}}{R^{2}}\approx
2.49\times 10^{-4}\;\mathrm{s}.
\end{equation}

\paragraph{Radar echo delay in semi-symmetric metric gravity.}

For the case of the semi-symmetric gravity black hole metric, by neglecting
the second order terms in $R_{g}^{2}$,  we have the metric (\ref{58}) $%
e^{\nu }=1-r_g/r+2\varepsilon \delta r/r_{g}$, and 
\begin{equation}
e^{-\nu }=1+\frac{r_{g}}{r}-2\varepsilon \delta \frac{r}{r_{g}},
\end{equation}
respectively. Then we obtain for the gravitational time delay the expression 
\begin{eqnarray}
\delta T_{RD} &=&\frac{1}{c}\int_{-l_{1}}^{l_{2}}{\left( \frac{r_{g}}{\sqrt{%
y^{2}+R^{2}}}-\frac{2\varepsilon \delta }{r_{g}}\sqrt{y^{2}+R^{2}}\right) dy}
\notag \\
&\approx &\frac{r_{g}}{c}\ln \frac{4l_{1}l_{2}}{R^{2}}-\frac{\varepsilon
\delta }{cr_{g}}\left( l_{2}^{2}+l_{1}^{2}+R^{2}\ln \frac{4l_{1}l_{2}}{R^{2}}%
\right)\nonumber\\
& =&\delta T_{RD}^{(Sch)}\left( 1+\Delta T_{RD}\right) ,  
\end{eqnarray}%
where 
\begin{equation}\label{isthismeasured?}
\Delta T_{RD}=-\varepsilon \frac{1}{R_{g}r_{g}}\frac{%
l_{2}^{2}+l_{1}^{2}+R^{2}\ln \left( 4l_{1}l_{2}/R^{2}\right) }{\ln \left(
4l_{1}l_{2}/R^{2}\right) }.
\end{equation}

Although present observations allow $\Delta T_{RD}$ to take either sign, we adopt $\varepsilon = -1$ to remain consistent with the other Solar System tests. Hence, for $R_{g}$ we find the constraint
\begin{equation}
R_{g}\approx \frac{l_{2}^{2}+l_{1}^{2}+R^{2}\ln \left(
4l_{1}l_{2}/R^{2}\right) }{r_{g}\ln \left( 4l_{1}l_{2}/R^{2}\right) }\frac{1%
}{\Delta T_{RD}}.
\end{equation}

For the time delay of the signals emitted on Earth, and which just touch the
Sun, one obtains $\Delta T_{RD}\approx (1.1\pm 1.2)\times 10^{-5}$ \cite{Ber}%
. Thus, for the case of the Earth-Mars-Sun system the radar echo delay test
gives for $R_{g}$ the numerical estimate $\;$

\begin{equation}
R_{g}\approx 2.074\times 10^{25}\;{\rm cm}.
\end{equation}

This value is very close to the value obtained from the study of the light deflection by the Sun.

\section{Horizon thermodynamics and linear perturbations}\label{sec:properties}
\subsection{{Thermodynamics of the black hole solutions}}
{We now investigate the thermodynamic properties of the exact black hole solutions obtained above. Following the standard approach adopted, for example, in \cite{PhysRevD.110.064012}, the horizons are determined by the condition
\begin{equation}
    g^{rr}=e^{-\lambda}=0.
\end{equation}
Using Eq.~\eqref{fin1}, this condition gives
\begin{equation}
    1- 2 \varepsilon \delta -\frac{r_g}{r} - \delta(\delta - 2 \varepsilon) \frac{r}{r_g} + \delta^{2} \frac{r^2}{r_g^2}=0,
\end{equation}
or, equivalently, after multiplication by $r$, a cubic equation for the horizon radius. Its positive roots are\footnote{{In the case $\varepsilon=1$, the solution $r=-R_{g}$ is discarded, as $R_{g}>0$ and $r$ cannot be negative.}}
\begin{enumerate}
    \item[$(i)$] $r_{1}=r_{g}$, for $\varepsilon=1$;
    \item[$(ii)$] $r_{2}=r_{g}$ and $r_{3}=R_{g}$, for $\varepsilon=-1$.
\end{enumerate}
Since the metric satisfies $e^{\nu}=e^{-\lambda}$, the Hawking temperature associated with a horizon $r_i$ can be computed from
\begin{equation}
\begin{aligned}
    T_{H}(r_{i})&=\left.\frac{\nu^{\prime}}{4 \pi} e^{\frac{\nu - \lambda}{2}} \right|_{r=r_{i}}.
\end{aligned}
\end{equation}
For $\varepsilon=1$, the black hole has a single physical horizon, $r_1=r_g$, with Hawking temperature
\begin{equation}
    T_{H}(r_1)=\frac{1}{4 \pi} \left(\frac{1}{r_g}+\frac{1}{R_g} \right).
\end{equation}
For $\varepsilon=-1$, one obtains two horizons. The black hole horizon at $r_2=r_g$ has temperature
\begin{equation}
T_{H}(r_2)
=
\frac{1}{4 \pi}
\left(
\frac{1}{r_g}
-
\frac{1}{R_g}
\right),
\end{equation}
whereas the outer horizon at $r_3=R_g$ has
\begin{equation}
T_{H}(r_3)=0 .
\end{equation}
}

{
The vanishing of the temperature at $r_3=R_g$ follows from the fact that this horizon is a double root of the metric function. Hence, it behaves as an extremal horizon. This is similar to cosmological horizons in Schwarzschild--de Sitter-type geometries, although in the present case the temperature vanishes exactly, rather than being merely extremely small.
}

{
The corresponding Bekenstein--Hawking entropies of the black hole horizons are
\begin{equation}
S(r_1)=S(r_2)
=
\frac{k_B A_i}{4l_P^2}
=
\frac{k_B 4\pi r_g^2}{4l_P^2}
=
\frac{\pi k_B r_g^2}{l_P^2},
\end{equation}
where $A_i=4\pi r_i^2$ is the horizon area and $l_P$ denotes the Planck length. %The heat capacitities are given by
%\begin{equation}
 %\begin{aligned}
  %  C_{H}(r_1)&=T_{H}(r_1) \frac{ \partial S(r_1)}{\partial T_{H}(r_1)}=%T_{H}(r_1) \frac{ \partial \left(\frac{\pi k_{B}}{ \left(l_{P} \left( 4 \pi T_{H}(r_1) - \frac{1}{R_g} \right) \right)^2}  \right)}{\partial T_{H}(r_1)}\\
   % -T_{H}(r_1) \frac{8 \pi^2 k_{B}}{l_P^2 \left( 4 \pi T_{H}(r_1) - \frac{1}{R_g} \right)^3},
%\end{aligned}
%\end{equation}
The heat capacity is defined as
\begin{equation}
C_H(r_i)
=
T_H(r_i)
\frac{\partial S(r_i)}{\partial T_H(r_i)} .
\end{equation}
%\begin{equation}
%\begin{aligned}
      %  C_{H}(r_2)&=T_{H}(r_2) \frac{ \partial S(r_2)}{\partial T_{H}(r_2)}=%T_{H}(r_2) \frac{ \partial \left(\frac{\pi k_{B}}{ \left(l_{P} \left( 4 \pi T_{H}(r_2) + \frac{1}{R_g} \right) \right)^2}  \right)}{\partial T_{H}(r_2)}\\
    %-T_{H}(r_2) \frac{8 \pi^2 k_{B}}{l_P^2 \left( 4 \pi T_{H}(r_2) + \frac{1}{R_g} \right)^3}.
%\end{aligned}
%\end{equation}
For the horizon $r_1=r_g$, corresponding to $\varepsilon=1$, this gives
\begin{equation}
\begin{aligned}
C_H(r_1)
&=
-T_H(r_1)
\frac{8\pi^2 k_B}
{l_P^2
\left(
4\pi T_H(r_1)-\frac{1}{R_g}
\right)^3}.
\end{aligned}
\end{equation}
Similarly, for the horizon $r_2=r_g$, corresponding to $\varepsilon=-1$, one obtains
\begin{equation}
\begin{aligned}
C_H(r_2)
&=
-T_H(r_2)
\frac{8\pi^2 k_B}
{l_P^2
\left(
4\pi T_H(r_2)+\frac{1}{R_g}
\right)^3}.
\end{aligned}
\end{equation}
Using the relations
\begin{equation}
4\pi T_H(r_1)-\frac{1}{R_g}
=
\frac{1}{r_1},
\qquad
4\pi T_H(r_2)+\frac{1}{R_g}
=
\frac{1}{r_2},
\end{equation}
the heat capacities can be written in the compact form
\begin{equation}
C_H(r_i)
=
-
\frac{8\pi^2 k_B}{l_P^2}
T_H(r_i) r_i^3 .
\end{equation}
}

{
Therefore, whenever the Hawking temperature is positive, the heat capacity is negative,
\begin{equation}
C_H(r_i)<0 .
\end{equation}
This is the same qualitative behavior as in the Schwarzschild case and indicates the usual thermodynamic instability of asymptotically flat black holes: as the black hole loses energy, its temperature increases.
}

%\textcolor{blue}{
%To see more clearly the effect of the extra scale $R_{g}$, we represent the heat capacities as
%\begin{equation}
%\begin{aligned}
  %  C_{H}(r_1)&=-2 S(r_1) \left( 1+ \frac{r_1}{R_g} \right), \\
   % C_{H}(r_2)&=-2S(r_2) \left( 1 - \frac{r_2}{R_g} \right).
%\end{aligned}
%\end{equation}
%}

%\textcolor{blue}{From the  above expressions, we see that the extra length scale $R_{g}$ only gives small corrections when the horizon size is much smaller than the cosmological scale, and in the limit $R_{g} \gg R_{i}$, one recovers approximately the Schwarzschild result $C_{H} \simeq -2 S$.
%}

%\textcolor{blue}{For the  cosmological horizon $r_{3}=R_{g}$, since $T_{H}(r_3)=0$, one obtains a vanishing heat capacity,
%\begin{equation}
   % C_{H}(r_3)=0,
%\end{equation}
%which can be interpreted as a thermally inert or cold horizon: although the horizon may still possess a finite area entropy, it does not radiate thermally and does not exchange heat in the usual Hawking sense.
%}
{
The effect of the additional length scale $R_g$ can be displayed more transparently by rewriting the heat capacities as
\begin{equation}
\begin{aligned}
C_H(r_1)
&=
-2S(r_1)
\left(
1+\frac{r_1}{R_g}
\right), \\
C_H(r_2)
&=
-2S(r_2)
\left(
1-\frac{r_2}{R_g}
\right).
\end{aligned}
\end{equation}
Thus, when the black hole horizon radius is much smaller than the new length scale, $r_i\ll R_g$, the torsional correction is small and one recovers the Schwarzschild result,
\begin{equation}
C_H \simeq -2S .
\end{equation}
}

{
For the outer horizon $r_3=R_g$, one has $T_H(r_3)=0$. Consequently, its heat capacity vanishes,
\begin{equation}
C_H(r_3)=0 .
\end{equation}
This horizon may therefore be interpreted as a cold horizon. Although it has a finite area and hence a finite Bekenstein--Hawking entropy, it does not radiate thermally in the usual Hawking sense.
}
\subsection{{Linear perturbations}}
{In this section we study scalar and vector perturbations on the static, spherically symmetric background introduced above. We denote
\begin{equation}
    A(r) \equiv \sqrt{\frac{g_{tt}(r)}{|g_{rr}(r)|}} ,
\end{equation}
and use a prime to indicate differentiation with respect to $r$.}
\par{{\textit{Scalar perturbations.}}} {The massless scalar perturbations propagating in a static spherically symmetric metric are governed by
\begin{equation}
\label{eq:KG_scalar_background}
    \frac{A}{r^2}\partial_r\left(r^2 A\partial_r\Phi\right)
    +\frac{g_{tt}}{r^2}\Delta_{\Omega}\Phi
    -\partial_t^2\Phi=0,
\end{equation}
where
\begin{equation}
    \Delta_{\Omega}
    =
    \frac{1}{\sin\theta}
    \partial_{\theta}
    \left(
    \sin\theta\,\partial_{\theta}
    \right)
    +
    \frac{1}{\sin^2\theta}
    \partial_{\phi}^2
\end{equation}
is the angular Laplacian on the unit two-sphere.  We decompose the scalar field as
\begin{equation}
\label{eq:scalar_decomposition}
    \Phi(t,r,\theta,\phi)
    =
    \frac{1}{r}
    \sum_{l,m}
    \Psi^s_{lm}(t,r)
    Y_{lm}(\theta,\phi),
\end{equation}
with
\begin{equation}
    Y_{lm}(\theta,\phi)
    =
    \sqrt{
    \frac{2l+1}{4\pi}
    \frac{(l-m)!}{(l+m)!}
    }
    P_l^m(\cos\theta)e^{im\phi}.
\end{equation}
Using
\begin{equation}
    \Delta_{\Omega}Y_{lm}
    =
    -l(l+1)Y_{lm},
\end{equation}
together with the orthogonality of the spherical harmonics, Eq.~\eqref{eq:KG_scalar_background} reduces, for each $(l,m)$ mode, to
\begin{equation}
\label{eq:scalar_radial_t}
    A^2\partial_r^2\Psi^s_{lm}
    +
    AA'\partial_r\Psi^s_{lm}
    -
    \partial_t^2\Psi^s_{lm}
    =
    \left[
    \frac{AA'}{r}
    +
    \frac{l(l+1)g_{tt}}{r^2}
    \right]
    \Psi^s_{lm}.
\end{equation}}

{It is convenient to introduce the tortoise coordinate
\begin{equation}
\label{eq:tortoise}
    \frac{dr_*}{dr}
    =
    \frac{1}{A}
    =
    \sqrt{\frac{
    |g_{rr}|}{g_{tt}}}.
\end{equation}
Since
\begin{equation}
    \partial_{r_*}^2
    =
    A^2\partial_r^2
    +
    AA'\partial_r,
\end{equation}
Eq.~\eqref{eq:scalar_radial_t} becomes
\begin{equation}
\label{eq:scalar_wave}
    \left(
    \partial_{r_*}^2
    -
    \partial_t^2
    -
    V_s(r)
    \right)
    \Psi^s_{lm}(t,r_*)=0,
\end{equation}
where the scalar effective potential is
\begin{equation}
\label{eq:scalar_potential_general}
    V_s(r)
    =
    \frac{l(l+1)g_{tt}(r)}{r^2}
    +
    \frac{1}{r}
    \sqrt{\frac{g_{tt}(r)}{|g_{rr}(r)|}}
    \frac{d}{dr}
    \left[
    \sqrt{\frac{g_{tt}(r)}{|g_{rr}(r)|}}
    \right].
\end{equation}
For the subclass satisfying $g_{rr}=1/g_{tt}$, this expression simplifies to
\begin{equation}
\label{eq:scalar_potential_f}
    V_s(r)
    =
    \frac{l(l+1)f(r)}{r^2}
    +
    \frac{f(r)f'(r)}{r},
    \qquad
    f(r)\equiv g_{tt}(r).
\end{equation}
Equivalently,
\begin{equation}
    V_s(r)
    =
    V_v(r)
    +
    \frac{f(r)f'(r)}{r},
    \qquad
    V_v(r)
    =
    \frac{l(l+1)f(r)}{r^2}.
\end{equation}
Assuming harmonic time dependence,
\begin{equation}
    \Psi^s_{lm}(t,r_*)
    =
    e^{-i\omega t}
    \Psi^s_{\omega lm}(r_*),
\end{equation}
we obtain the Schrödinger-like equation
\begin{equation}
\label{eq:scalar_schrodinger}
    \frac{d^2\Psi^s_{\omega lm}}{dr_*^2}
    +
    \left[
    \omega^2
    -
    V_s(r)
    \right]
    \Psi^s_{\omega lm}=0.
\end{equation}}

{For the black-hole solutions considered here, we introduce
\begin{equation}
    \delta=\frac{r_g}{R_g},
    \qquad
    x=\frac{r}{r_g},
    \qquad
    L=l(l+1).
\end{equation}
The metric function may then be written as
\begin{equation}
\label{eq:f_epsilon}
    f_{\varepsilon}(x)
    =
    \frac{
    (x-1)
    \left(
    1+2\varepsilon\delta x+\delta^2x^2
    \right)
    }
    {
    (1+\varepsilon\delta)x
    }
    =
    \frac{
    (x-1)
    \left(
    1+\varepsilon\delta x
    \right)^2
    }
    {
    (1+\varepsilon\delta)x
    }.
\end{equation}
Thus the scalar potential becomes
\begin{equation}
\label{eq:scalar_potential_dimensionless}
    V_s(r)
    =
    \frac{1}{r_g^2}
    \left[
    \frac{L f_{\varepsilon}(x)}{x^2}
    +
    \frac{f_{\varepsilon}(x)}{x}
    \frac{d f_{\varepsilon}}{dx}
    \right].
\end{equation}
It is useful to define the dimensionless scalar potential by
\begin{equation}
    V_s(r)=\frac{1}{r_g^2}U_s^{\varepsilon}(x),
\end{equation}
where
\begin{equation}
\label{eq:scalar_U}
    U_s^{\varepsilon}(x)
    =
    \frac{L f_{\varepsilon}(x)}{x^2}
    +
    \frac{f_{\varepsilon}(x)}{x}
    \frac{df_{\varepsilon}}{dx}.
\end{equation}
Introducing the notation $
    q_{\varepsilon}\equiv \varepsilon\delta$, we have
\begin{equation}
\label{eq:scalar_U_explicit}
\begin{aligned}
    U_s^{\varepsilon}(x)
    =
    \frac{
    (x-1)(1+q_{\varepsilon}x)^2
    }
    {
    (1+q_{\varepsilon})^2x^4
    } \Bigg[
    &L(1+q_{\varepsilon})x+
    2q_{\varepsilon}^2x^3\\
    &-
    q_{\varepsilon}^2x^2
    +
    2q_{\varepsilon}x^2
    +
    1 \Bigg].
\end{aligned}
\end{equation}
}
{
The maximum of the scalar potential is determined by
\begin{equation}
\label{eq:scalar_peak_condition}
    \left.
    \frac{dU_s^{\varepsilon}}{dx}
    \right|_{x=x_s}
    =
    0 .
\end{equation}
Equivalently, the physical peak \(x_s\) satisfies
\begin{equation}
\label{eq:scalar_peak_polynomial}
\begin{aligned}
0
=&\;
L(1+q_{\varepsilon})x_s
\left[
(q_{\varepsilon}-2)x_s+3
\right]
\\
&+
(1+q_{\varepsilon}x_s)
\Big[
4q_{\varepsilon}^2x_s^4
-
3q_{\varepsilon}^2x_s^3
-
2q_{\varepsilon}x_s^3
+
6q_{\varepsilon}x_s^2
\\
&\hspace{3.2cm}
-
2q_{\varepsilon}x_s
-
3x_s
+
4
\Big] .
\end{aligned}
\end{equation}
}

{
Using the first-order WKB approximation,
\begin{equation}
    \omega^2
    \simeq
    V_0
    -
    i
    \left(
    n+\frac{1}{2}
    \right)
    \sqrt{-2V_0''},
\end{equation}
where \(V_0\) and \(V_0''\) are evaluated at the maximum of the potential, the scalar quasinormal spectrum is
\begin{equation}
\label{eq:omega_scalar_squared_general}
    \left(
    \omega_s^{\varepsilon}
    \right)^2
    \simeq
    \frac{1}{r_g^2}
    \left[
    U_s^{\varepsilon}(x_s)
    -
    i
    \left(
    n+\frac{1}{2}
    \right)
    \sqrt{
    -2K_s^{\varepsilon}(x_s)
    }
    \right],
\end{equation}
with
\begin{equation}
\label{eq:scalar_K_definition}
    K_s^{\varepsilon}(x_s)
    =
    \left.
    f_{\varepsilon}^2(x)
    \frac{d^2U_s^{\varepsilon}}{dx^2}
    \right|_{x=x_s}.
\end{equation}
Therefore,
\begin{equation}
\label{eq:omega_scalar_general}
    \omega_s^{\varepsilon}
    \simeq
    \frac{1}{r_g}
    \sqrt{
    U_s^{\varepsilon}(x_s)
    -
    i
    \left(
    n+\frac{1}{2}
    \right)
    \sqrt{
    -2K_s^{\varepsilon}(x_s)
    }
    },
\end{equation}
where the branch with \(\operatorname{Im}(\omega_s^{\varepsilon})<0\) is chosen.
}

{From Eq. \eqref{eq:scalar_peak_polynomial} it can be seen that the quasinormal modes cannot be determined analytically, as the maximum point $x_{s}$ is the solution of a quintic equation. For this reason, we move on to the vector perturbations, which can be computed analytically, and leave for future work the explicit numerical determination of the scalar modes, alongside a comparison with data.}

\par{{\textit{Vector perturbations.}}} 
{Vector perturbations are described by the antisymmetric field-strength tensor $F_{\mu\nu}$. Following the standard separation of Maxwell perturbations on a static and spherically symmetric background, we introduce
\begin{equation}
    \mathcal{F}
    =
    F_{t\phi}\sin\theta .
\end{equation}
The corresponding master equation is
\begin{equation}
\label{eq:vector_master_background}
\begin{aligned}
    \partial_r
    \left[
    A\,
    \partial_r
    \left(
    r\sqrt{g_{tt}}\,\mathcal{F}
    \right)
    \right]
    &+
    \frac{g_{tt}\sqrt{|g_{rr}|}}{r}
    \sin\theta\,
    \partial_{\theta}
    \left(
\frac{\partial_{\theta}\mathcal{F}}{\sin\theta}
    \right)\\
    &
    -
    r\sqrt{|g_{rr}|}\,
    \partial_t^2\mathcal{F}
    =
    0 .
\end{aligned}
\end{equation}
}

{
We separate variables according to
\begin{equation}
\mathcal{F}(t,r,\theta, \phi) = \sum_{l} e^{-i \omega t} \Psi^{v}_{\omega l}(r) \Theta_{l}(\theta),
\end{equation}
where the angular functions satisfy
\begin{equation}
    \sin\theta
    \frac{d}{d\theta}
    \left(
    \frac{1}{\sin\theta}
    \frac{d\Theta_l}{d\theta}
    \right)
    =
    -l(l+1)\Theta_l .
\end{equation}
For each $l$ mode, one obtains
\begin{equation}
\label{eq:vector_radial_original}
\begin{aligned}
    \partial_r
    \left[
    A\,
    \partial_r
    \left(
    r\sqrt{g_{tt}}
    \Psi^v_{\omega l}
    \right)
    \right]
    &+
    \omega^2 r\sqrt{|g_{rr}|}
    \Psi^v_{\omega l}\\
    &-
    \frac{g_{tt}\sqrt{|g_{rr}|}}{r}
    l(l+1)
    \Psi^v_{\omega l}
    =
    0 .
\end{aligned}
\end{equation}
Defining the master variable
\begin{equation}
    \Phi^v_{\omega l}(r)
    =
    r\sqrt{g_{tt}}
    \Psi^v_{\omega l}(r),
\end{equation}
and using the tortoise coordinate \eqref{eq:tortoise}, we have
\begin{equation}
    \partial_r
    \left(
    A\partial_r\Phi^v_{\omega l}
    \right)
    =
    \sqrt{\frac{|g_{rr}|}{g_{tt}}}\,
    \frac{d^2\Phi^v_{\omega l}}{dr_*^2}.
\end{equation}
Equation \eqref{eq:vector_radial_original} therefore becomes
\begin{equation}
\label{eq:vector_schrodinger}
    \frac{d^2\Phi^v_{\omega l}}{dr_*^2}
    +
    \left[
    \omega^2
    -
    V_v(r)
    \right]
    \Phi^v_{\omega l}
    =
    0,
\end{equation}
with the vector effective potential
\begin{equation}
\label{eq:vector_potential_general}
    V_v(r)
    =
    \frac{l(l+1)g_{tt}(r)}{r^2}.
\end{equation}
%where $\Psi^{v}_{\omega l}(r)$ denotes radial part  and the angular part $\Theta_{l}(\theta)$ satisfies
%\begin{equation}
   % \sin \theta \frac{d}{d \theta} \left(\frac{1}{\sin \theta} \frac{d \Theta_{l}(\theta)}{d \theta} \right)=-l(l+1) \Theta_{l}(\theta),
%\end{equation}
%we obtain for each $l$ mode
%\begin{equation}
%\begin{aligned}
 %   &\left[\sqrt{g_{tt} g_{rr}^{-1}} \left( r \sqrt{g_{tt}} \Psi_{\omega l}^{v} \right)_{,r}\right]_{,r} +\omega^2 r \sqrt{g_{rr}} \Psi_{\omega l}^{v}\\
   % &- \frac{g_{tt} \sqrt{g_{rr}}}{r}l(l+1) \Psi_{\omega l}^{v}=0.
%\end{aligned}
%\end{equation}
%Working in the previously introduced tortoise coordinates \eqref{eq:tortoise}, we redefine the radial perturbations through
%\begin{equation}
    %\Phi^{v}_{\omega l}(r)=r \sqrt{g_{tt}} \Psi^{v}_{\omega l}(r).
%\end{equation}
%This redefinition implies
%\begin{equation}
   % \left[\sqrt{g_{tt} g_{rr}^{-1}} \left( r \sqrt{g_{tt}} \Psi_{\omega l}^{v} \right)_{,r}\right]_{,r}=\frac{d \Phi^{v}_{\omega l}(r_*)}{d r_{*}},
%\end{equation}
%so that the radial equation becomes
%\begin{equation}
    %\sqrt{\frac{g_{rr}}{g_{tt}}} \frac{ d^2 \Phi^{v}_{\omega l}(r_*)}{ dr_*^2} + \sqrt{\frac{g_{rr}}{g_{tt}}}{}\left[\omega^2 - \frac{g_{tt}}{r^2}l(l+1) \right] \Phi_{\omega l}^{v}(r_*)=0.
%\end{equation}
%Equivalently, it can be expressed as
%\begin{equation}
  %  \frac{d^2 \Phi_{\omega l}^{v}(r_*)}{dr_*^2}+\omega^2 \Phi_{\omega l}^{v}(r_*)=V_{v}(r) \Phi_{\omega l}^{v}(r_*),
%\end{equation}
%where $V_{v}(r)$ denotes potential for vector perturbations
%\begin{equation}
   % V_{v}(r)=g_{tt}(r) \frac{l(l+1)}{r^2}.
%\end{equation}
For the present black hole metric, this gives
\begin{equation}
\begin{aligned}
    V_{v}(r)=\frac{l(l+1)}{r^2 \left(1 + \varepsilon \frac{r_g}{R_g} \right)} \Bigg[&1-2
\varepsilon \frac{r_g}{R_g}-\frac{r_g}{r}\\
&-\frac{r_g - 2 \varepsilon R_g}{
R_g^2}r+\frac{r^2}{R_g^2} \Bigg].
\end{aligned}
\end{equation}
Thus the terms controlled by $R_g$ encode the large-distance modification of the effective barrier. In the limit $R_g\gg r_g$, one recovers the Schwarzschild result
\begin{equation}
    V_v(r)
    \simeq
    \frac{l(l+1)}{r^2}
    \left(
    1-\frac{r_g}{r}
    \right).
\end{equation}
}

{In terms of the dimensionless variables introduced above,
\begin{equation}
\label{eq:vector_U}
    V_v(r)
    =
    \frac{L}{r_g^2}
    U_{\varepsilon}(x),
    \qquad
    U_{\varepsilon}(x)
    =
    \frac{
    (x-1)
    \left(
    1+\varepsilon\delta x
    \right)^2
    }
    {
    (1+\varepsilon\delta)x^3
    }.
\end{equation}
%\textcolor{blue}{
%The quasinormal modes can be estimated using the first-order WKB formula
%\begin{equation}
 %   \omega^{2} \simeq V_{0}-i \left( n+ \frac{1}{2}\right) \sqrt{-2 V_{0}''},
%\end{equation}
%where $V_{0}$ is the maximum of the potential $V_{l}$ and $V_{0}^{''}$ is the second derivative with respect to the tortoise coordinate $r_{*}$, evaluated at the maximum.
%}
%\textcolor{blue}{Introducing dimensionless variables
%\begin{equation}
 %   \delta=\frac{r_g}{R_g}, \; \; x=\frac{r}{r_g}, \; \; L=l(l+1),
%\end{equation}
%the potential can be written as
%\begin{equation}
   % V_{v}(r)=\frac{L}{r_g^2} U_{\epsilon}(x),
%\end{equation}
%where we defined
%\begin{equation}
   % U_{\epsilon}(x)=\frac{(x-1)(1+2 \varepsilon \delta x + \delta^2 x^2)}{(1+\varepsilon \delta) x^3}.
%\end{equation}
%\textcolor{blue}{
%To find the maximum, we set the derivative of $U_{\varepsilon}(x)$ to zero, obtaining 
The maximum of the potential is determined by
\begin{equation}
    \frac{d U_{\varepsilon}}{dx}=0,
\end{equation}
which gives
\begin{equation}
   (\delta^2-2 \varepsilon \delta)x^2 + \left( 4 \varepsilon \delta -2\right)x +3=0.
\end{equation}
The root continuously connected to the Schwarzschild value $x_{0}=\frac{3}{2}$ is\footnote{{Here, we used that $\varepsilon=\pm 1$ and $0 \leq \delta < 1$. It is also interesting to note that for $\varepsilon=-1$, one root is $x=\frac{1}{\delta}$, the cosmological horizon, which is not the peak. Hence, our selected root is the physical maximum away from the horizon.}}
\begin{equation}
\label{eq:x0_general}
    x_0
    =
    \frac{3}{2-\varepsilon\delta},
    \qquad
    r_0
    =
    r_g x_0 .
\end{equation}
%\begin{equation}
   % x_{0}=\frac{2 \varepsilon \delta + \sqrt{1+ 2\varepsilon \delta+(4 \varepsilon^2-3)\delta^2}-1}{\delta(2 \varepsilon - \delta)},
%\end{equation}
%so that $r_{0}=r_{g} x_{0}$.
At this point,
\begin{equation}
\begin{aligned}
\label{eq:UK_peak}
    U_{\varepsilon}(x_0)
    &=
    \frac{4(1+\varepsilon\delta)^2}{27},\\
    K_{\varepsilon}(x_0)
    &\equiv
    \left[
    f_{\varepsilon}^2
    \frac{d^2U_{\varepsilon}}{dx^2}
    \right]_{x=x_0}
    =
    -\frac{32(1+\varepsilon\delta)^4}{729}.
\end{aligned}
\end{equation}
}

%\textcolor{blue}{The WKB quasinormal spectrum can therefore be written as
%\begin{equation}
   % \omega^{2}_{v} \simeq \frac{1}{r_g^2} \left[ L U_{\varepsilon}(x_0) - i \left(n+\frac{1}{2} \right) \sqrt{-2L K_{\varepsilon}(x_0)} \right],
%\end{equation}
%where the subscript $v$ stands for vectorial, and
%\begin{equation}
  %  K_{\varepsilon}(x_0)=\left[g_{tt}^2 \frac{d^2 U_{\varepsilon}}{dx^2} \right]_{x=x_{0}}.
%\end{equation}
%}
{
The quasinormal modes can be estimated with the first-order WKB formula
\begin{equation}
\label{eq:first_order_wkb}
    \omega^2
    \simeq
    V_0
    -
    i
    \left(
    n+\frac{1}{2}
    \right)
    \sqrt{-2V_0''},
\end{equation}
where $V_0$ is the maximum of the effective potential, and $V_0''$ denotes the second derivative with respect to $r_*$, evaluated at the maximum. Using Eqs.~\eqref{eq:vector_U}--\eqref{eq:UK_peak}, the vector spectrum becomes
\begin{equation}
\label{eq:omega_vector_squared}
    \left(
    \omega_v^{\varepsilon}
    \right)^2
    \simeq
    \frac{4(1+\varepsilon\delta)^2}{27r_g^2}
    \left[
    L
    -
    2i
    \left(
    n+\frac{1}{2}
    \right)
    \sqrt{L}
    \right].
\end{equation}
Hence
\begin{equation}
\label{eq:omega_vector_exact_wkb_form}
    \omega_v^{\varepsilon}
    \simeq
    \frac{2(1+\varepsilon\delta)}
    {3\sqrt{3}\,r_g}
    \sqrt{
    L
    -
    2i
    \left(
    n+\frac{1}{2}
    \right)
    \sqrt{L}
    }.
\end{equation}
}

{
In the eikonal limit $l \gg 1$,
this reduces to\footnote{{Here, we used that $
    \sqrt{L-2 i \left( n+ \frac{1}{2} \right) \sqrt{L}} \simeq \sqrt{L} - i \left( n +\frac{1}{2} \right)$.}}
\begin{equation}
\label{eq:omega_vector_eikonal}
    \omega_v^{\varepsilon}
    \simeq
    \frac{2}{3\sqrt{3}\,r_g}
    \left(
    1+\varepsilon\frac{r_g}{R_g}
    \right)
    \left[
    \sqrt{l(l+1)}
    -
    i
    \left(
    n+\frac{1}{2}
    \right)
    \right].
\end{equation}
}
%\begin{equation}
   % \omega_{v}^{+} \simeq \frac{2}{3 \sqrt 3 r_{g}} \left( 1+\frac{r_g}{R_g}\right) \left[ \sqrt{l(l+1)}-i \left(n+\frac{1}{2} \right)\right].
%\end{equation}
%}
%\textcolor{blue}{In an exact similar manner, for $\varepsilon=-1$, one finds
%\begin{equation}
  %  \omega_{v}^{-} \simeq \frac{2}{3 \sqrt{3} r_{g}} \left( 1 - \frac{r_g}{R_g} \right) \left[ \sqrt{l(l+1)}-i \left(n+\frac{1}{2} \right)\right],
%\end{equation}
%so that the quasinormal frequencies %can be compactly written as
%\begin{equation}
  %  \omega_{v}^{\epsilon} \simeq \frac{2}{3 \sqrt 3 r_{g}} \left( 1 + \varepsilon
 %\frac{r_g}{R_g} \right) \left[ \sqrt{l(l+1)}-i \left(n+\frac{1}{2} \right)\right].\end{equation}
 %For $\varepsilon=+1$, the $R_{g}$-dependent correction raises the vector potential barrier and increases both the oscillation frequency and damping rate of the quasinormal modes. For $\varepsilon=-1$, the correction lowers the barrier, and the presence of the cold cosmological horizon at $r=R_{g}$ suppresses the QNM spectrum by the factor $1-\frac{r_g}{R_g}$. In the limit $R_{g} \to r_{g},$ the vector modes become increasingly soft and long-lived, while for $R_{g} \gg r_{g}$ the standard Schwarzschild-like spectrum is recovered.

{For $\varepsilon=+1$, the correction proportional to $R_g^{-1}$ raises the vector potential barrier and increases both the oscillation frequency and the damping rate. For $\varepsilon=-1$, the barrier is lowered, and the cold cosmological horizon at $r=R_g$ suppresses the spectrum by the factor $1-r_g/R_g$. The Schwarzschild-like result is recovered for $R_g\gg r_g$.}

%\textcolor{blue}{One can also observe that the vector sector is linearly stable. Indeed, for $l\geq 1$, the effective potential is non-negative in the physical region. For the $\varepsilon=-1$ branch, corresponding to the cold cosmological horizon at $r=R_g$, the potential reduces to
%\begin{equation}
%V_v(r)=
%\frac{l(l+1)}{r^2\left(1-r_g/R_g\right)}
%\left(1-\frac{r_g}{r}\right)
%\left(1-\frac{r}{R_g}\right)^2 .
%\end{equation}
%This potential is positive for $r_g<r<R_g$ and vanishes only at the horizons. Thus, there is no negative potential well capable of supporting unstable bound states. The WKB quasinormal spectrum has $\operatorname{Im}(\omega)<0$, so the perturbations decay in time. For the $\varepsilon=+1$ branch, the potential is also positive outside the black-hole horizon and the imaginary part of the quasinormal frequencies remains negative. Therefore, in both branches, vector perturbations are damped and the background is stable at the linear level in the vector sector.}

{Finally, the vector sector is linearly stable in the physical region. For $l\geq1$, the effective potential is non-negative. In particular, for the $\varepsilon=-1$ branch,
\begin{equation}
\label{eq:vector_potential_minus_branch}
    V_v(r)
    =
    \frac{l(l+1)}
    {r^2(1-r_g/R_g)}
    \left(
    1-\frac{r_g}{r}
    \right)
    \left(
    1-\frac{r}{R_g}
    \right)^2,
\end{equation}
which is positive for $r_g<r<R_g$ and vanishes only at the horizons. Therefore, no negative potential well is present in this sector. Together with $\operatorname{Im}(\omega)<0$ in Eq.~\eqref{eq:omega_vector_eikonal}, this shows that vector perturbations decay in time.}

\section{Discussions and final remarks}\label{sect4}

In this paper, we have derived two exact analytical solutions of the vacuum
field equations in semi-symmetric metric gravity, assuming that the torsion
vector has only a radial component and imposing $\nu +\lambda =0$. We obtained  two exact solutions with the metric given by a function of the form
$e^\nu=C(r/R_g)^2\left(1\pm R_g/r\right)^2\left(1-r_g/r\right)$, where $R_g$ and $r_g$ are constants that do not depend on the initial conditions at
infinity. 

The classical tests of general relativity provide an efficient method for
constraining the parameter space of black hole solutions and probing the
physical properties of the corresponding spacetimes. Using the exact
spherically symmetric solution of the semi-symmetric metric gravity field
equations, we have shown that Solar System observations can be used to test
this theory. A general formalism was adopted that applies to any spherically
symmetric metric, allowing direct comparison of the predictions of
semi-symmetric metric gravity with both observations and standard general
relativity. Applying the classical tests, we have obtained explicit
constraints on the metric parameter $R_{g}$. The perihelion precession of
Mercury gives $R_{g}\approx 5\times 10^{26}$ cm, while light deflection by
the Sun and radar echo delay in the Earth-Mars system yield $R_{g}\approx
\left( 1.69- 2\right) $$\times 10^{25}$ cm. These correspond to the estimates 
\begin{equation*}
\frac{1}{R_{g}^{2}}\simeq 4\times 10^{-54}\;\mathrm{cm}^{-2},\quad \frac{1}{%
R_{g}^{2}}\approx 2.5\times 10^{-51}\;\mathrm{cm}^{-2}
\end{equation*}%
respectively. Although these values differ from the observed cosmological
constant $\Lambda \approx 10^{-56}\;\mathrm{cm}^{-2}$, they suggest
interpreting $R_{g}$ as related to $\Lambda $ via $R_{g}\propto 1/\sqrt{%
\Lambda }$. Nevertheless, it cannot be excluded that $R_{g}$ represents a
distinct length scale associated with torsional effects. 

{ If $R_g$ is interpreted as a cosmological length scale, the theory may also have implications for the early Universe. In an FLRW spacetime, homogeneity and isotropy restrict the torsion vector to a temporal component $\pi_\mu=(\omega(t),0,0,0)$, and the corresponding modified Friedmann equations admit de Sitter-type solutions  without introducing additional scalar fields \cite{32}. Vectorial torsion may therefore generate the accelerated expansion required for inflation. A complete inflationary scenario would, however, also require a mechanism for ending inflation, reheating, and a detailed analysis of scalar and tensor perturbations. These issues lie beyond the scope of the present work, and will be investigated in future. }

{The horizon thermodynamics allows to describe the effects of this possibly distinct length scale. For the $\varepsilon=1$ branch, the solution contains a single black-hole horizon at $r=r_g$, while for $\varepsilon=-1$ an additional outer horizon appears at $r=R_g$. The latter has vanishing Hawking temperature, since it corresponds to a double root of the metric function, and can therefore be interpreted as a cold horizon. The black hole horizons have negative heat capacity whenever their Hawking temperature is positive, showing the same qualitative thermodynamic instability as the Schwarzschild black hole.}

In the expression of the metric the term $r^2/R_g^2$ is always positive, and
it is thus independent of the sign of $\varepsilon$, the sign of the term
proportional to $r$ depends on the sign of $\varepsilon$, and it can be generally
approximated as $\pm 2r/R_g$. Since $R_g$ has very high values, the effects
of this term are relevant on length scales close to $R_g$, of the order of
megaparsecs. Hence, we expect that this term may play an important role in
giving physical explanation for the (yet unknown) dynamics of massive test
particles gravitating on circular orbits around the galactic centers, which
is usually interpreted by postulating the existence of dark matter.
Moreover, this term may become dominant for the description of the physical
processes taking place in the galactic clusters, astrophysical systems
consisting of thousands of galaxies in virial equilibrium and having
diameters in the range of 1-5 Mpc.

At distances of the order of Mpc's, corresponding to galactic cluster
scales, the torsion vector can be approximated as $\psi (r)\approx
\left(r/R_g^2\right)\left(1\pm R_g/r\right)$, since for a galaxy cluster
with mass $10^{15}M_{\odot}$ and radius 2 Mpc, the gravitational radius is
of the order of $r_g\approx 4.3\times 10^{16}$ cm, and hence $r\gg r_g$, and 
$R_g\gg r_g$, respectively. Therefore, torsion may be present at least at
the level of the clusters of galaxies, where it can exert a significant
effect on the overall dynamics, and even play the role of dark matter in
explaining the virial deficit without introducing dark matter.

The exact vacuum solution of semi-symmetric metric gravity naturally allows
the introduction of an effective geometric mass $M_{eff}$, and of an
associated effective density $\rho_{eff}$, which we interpret as describing
"dark matter", represented by a geometric effect. Based on this
interpretation, it is possible to observationally constrain $M_{eff}$ and $%
\rho_{eff}$ by applying a method based on the Jeans equation \cite{Jeans}.
We consider a simplified approach to the galactic dynamics by assuming that
each galaxy contains a single type of stellar population, pressure
supported, and which is in virial equilibrium with the gravitational field.
Moreover, the presence of the torsion at the galactic scale gives a
significant gravitational contribution, leading to the modification of the
space-time structure.

In static spherical symmetry, the effective geometric mass profile $M_{eff}$%
, assumed to represent, from an astrophysical point of view, the mass
profile of the dark matter halo, is related to the moments of the stellar
distribution function via the Jeans equation \cite{Jeans} 
\begin{equation}
\frac{d}{dr}\left[\rho_s(r)\left<v_r^2\right>\right]+\frac{2\beta
(r)\rho_s(r)\left<v_r^2\right>}{r}=-\frac{G\rho_s(r)M_{eff}(r)}{r^2},
\end{equation}
where $\beta (r)=1-\left<v_\theta^2\right>/\left<v_r^2\right>$ is the
orbital anisotropy of the stellar component, and $\rho_s(r)$, $%
\left<v_r^2\right>$, and $\left<v_\theta ^2\right>$ are the
three-dimensional stellar density, and the radial and tangential velocity
dispersion of the stars, respectively. Assuming that the anisotropy is
constant, the solution of the Jeans equation is given by 
\begin{equation}  \label{88}
\rho_s(r)\left<v_r^2\right>=Gr^{-2\beta}\int_r^{\infty}{x^{2(1-\beta)}%
\rho_s(x)M_{eff}(x)dx}.
\end{equation}

In the following we will adopt for the stellar density the expression based
on the Plummer profile, which is given by 
\begin{equation}
\rho_s(r)=\frac{3L_0}{4\pi r_h^3\left(1+r^2/r_h^2\right)^{5/2}},
\end{equation}
where $L_0$ is the total galactic luminosity, and $r_h$ denotes the
projected half-light radius, which is the radius of the cylinder containing
half of the total galactic luminosity. We also assume that the stellar
component is anisotropically distributed, and for simplicity we take $\beta
=1/2$. Hence, after substituting the expression of the stellar density and
of the effective mass into Eq.~(\ref{88}), we obtain the expression of the
radial velocity dispersion of the stars as 
\begin{eqnarray}  \label{90}
\left\langle v_{r}^{2}\right\rangle & =&\frac{G}{r\rho _{s}(r)}\Bigg\{ \frac{%
L_{0}}{4\pi R_{g}^{2}\left( 1+\frac{x^{2}}{r_{h}^{2}}\right) ^{3/2}}\Bigg[ %
-2r_{h}^{4}(r_{g}-2R_{g})  \notag \\
&&+2x^{3}\left( r_{g}R_{g}+2r_{h}^{2}\right)
-3r_{h}^{2}x^{2}(r_{g}-2R_{g})+3r_{h}^{4}x  \notag \\
&&-3r_{h}^{2}\left( r_{h}^{2}+\left.x^{2}\right) ^{3/2}\log \left( \sqrt{%
r_{h}^{2}+x^{2}}+x\right) \Bigg] \Bigg\}\right|_r^{\infty}.  \notag \\
\end{eqnarray}

Since Eq.~(\ref{90}) diverges at infinity, the upper limit must be replaced
by the upper bound of a cutoff radius, which can be chosen, for example, as $%
\Lambda$, or $R_g$. Hence, the numerical value of $R_g$ as well as the
expression of the torsion vector, can be obtained from the study of the
stellar velocity dispersion in the galactic clusters.

{Besides these astrophysical implications, the perturbative behavior of the solutions provides another consistency test of the obtained solutions. In the vector sector, the effective potential is non-negative in the physical region, while the imaginary part of the WKB quasinormal frequencies is negative. Hence, vector perturbations decay in time, indicating linear stability of the background in this sector. The scalar perturbation spectrum is more involved, since the position of the maximum of the effective potential is determined by a quintic equation. A detailed numerical investigation of scalar perturbations is therefore left for future work.}

The analytical solutions are very useful tools in the study of the physical
properties of black holes, including the investigation of the dynamics and
motion of massive particles around them. The black hole solutions can also
be used for the investigation of the electromagnetic properties of the thin
accretion disks that form around them. Moreover, the exact solution may also
help in discriminating between semi-symmetric black holes and standard
general relativistic black holes, and for obtaining observational
constraints on the effects and presence of the torsion in astrophysical
systems. The astrophysical implications of the present exact solution of the
semi-symmetric gravity theory will be investigated in detail in a further
study.

\section*{Acknowledgements}

{The authors would like to express their gratitude to the anonymous referee for the constructive and substantial comments. }L.Cs. would like to thank Collegium Talentum and the StarUBB research
scholarship for the support offered during the preparation of this
manuscript.

\end{document}